\documentclass[12pt]{iopart}

\usepackage{amssymb,dsfont,braket}
\usepackage[mathscr]{euscript}
\usepackage{graphicx}
\usepackage{epstopdf}
\usepackage[breaklinks]{hyperref}

\newcommand{\eqref}[1]{{(\ref{#1})}}

\newcommand{\U}[1]{{\mathrm{U}(#1)}}

\usepackage[usenames,dvipsnames]{color}
\definecolor{purple}{rgb}{0.5,0,0.5}

\begin{document}

\title{Flux insertion, entanglement, and quantized responses}
\author{Michael P. Zaletel}
\address{Department of Physics, University of California, Berkeley, California 94720, USA}
\author{Roger S. K. Mong}
\address{Department of Physics, California Institute of Technology, Pasadena, California 91125, USA}
\author{Frank Pollmann}
\address{Max-Planck-Institut f\"ur Physik komplexer Systeme, 01187 Dresden, Germany}

\begin{abstract}
We discuss the construction of a ground state wavefunction on a finite ring given the ground state on a long chain.
%, via entanglement cuts.
%We consider a system with symmetry group $G$ and discuss why the entanglement spectrum together with its quantum numbers encode several quantized response functions.
In the presence of local symmetries, we can obtain the ground state with arbitrary flux inserted through the ring.
A key ingredient are the quantum numbers of the entanglement spectrum.
This method allows us to characterize phases by measuring quantized responses, such as the Hall conductance, using data contained  in the entanglement spectrum.
%If the symmetry is a continuous symmetry, we show how to efficiently calculate a Berry phase that characterizes symmetry-protected topological phases directly from the entanglement spectrum.
As concrete examples, we show how the Berry phase allows us to map out the phase diagram of a spin-1 model and calculate the Hall conductivity of a quantum Hall system.  
\end{abstract}

\maketitle
\section{Introduction}
The study of entanglement has been shown to be very useful for the characterization of quantum states -- in particular for the understanding of topological properties of quantum states \cite{Wen-1990}.
Intrinsic topological order can for example be detected by an additive correction to the area law of the entanglement entropy \cite{Levin-2006,Kitaev-2006}.
In addition to the usefulness in classifying and detecting topological phases,  efficient numerical algorithms, like the density matrix renormalization group (DMRG)\cite{White-1992} method, are based on properties of the entanglement in ground-state wavefunction. 
Reversely, the DMRG method  gives us a direct access to the entanglement of the wave function.

Beyond the entanglement entropy, we can consider the full spectrum of the reduced density matrix $\rho_A$ for a subregion $A$, which we write as $\rho_A = e^{-H_E}$.  The spectrum of $H_E$ is the `entanglement spectrum'.
In certain circumstances the low lying eigenvalues of $H_E$ appear to reflect the energy spectrum of the edge states, and hence are a fingerprint of the topological order \cite{Kitaev-2006, Li-2008, Papic-2011}.
However, the correspondence between the low lying spectrum and the topological order cannot be taken too literally, since the low lying entanglement spectrum may undergo  a phase transition while the bulk phase remains unchanged \cite{Chandran-2014}.
It is worth clarifying which properties of $H_E$ are \emph{universal} properties of the bulk phase.

In this paper we review a class of properties of $H_E$ which are universal in the presence of symmetries.
The basic result is that the spectrum of $H_E$ \emph{combined} with the actions of the symmetries on its eigenstates is sufficient to predict the response of the phase to flux insertion, linking the entanglement spectrum to a well known class of quantized response functions such as polarization, Hall conductance, and the modular $T$-transformation \cite{Wen-1990} which encodes the statistics of anyonic quasiparticles. While these responses have previously been discussed elsewhere, \cite{Zak-1989, RestaVanderbilt, Laughlin-1981, NiuThoulessWu-1985, Alexandradinata-2011, Keski-Vakkuri-1993, WenMonodromy-2013} we find it useful to discuss them in the common language of entanglement, both in order to better understand the universality of $H_E$ and for the practical benefit that these responses can then be measured from entanglement information readily available using DMRG.

To link entanglement to flux insertion, we first show that given the ground state and entanglement spectrum of an infinite chain (cylinder) with symmetry group $G$, we can obtain the ground state on a ring (torus) with arbitrary flux inserted with respect to a symmetry $g \in G$.
This method can be useful for actual numerical simulations because DMRG simulations of systems with periodic boundary conditions are generically harder than with open (or infinite) boundary conditions. 
Various Berry phases and quantized responses exist for a ring (torus) with flux, and using this `flux insertion trick' we can then  extract the responses from the bulk entanglement spectrum of the infinite chain.
This paper is organized as follows.
We begin by introducing the flux insertion trick in Sec.~\ref{sec:flux} and discuss a spin-1 chain as concrete example.
In Sec.~\ref{sec:examples}, we give  applications to charge polarization (the `Zak phase') and the topological $T$ transformation (Sec.~\ref{sec:BerryU1}),   1D SPT phases (Sec.~\ref{sec:1d_spt}), and finally the Hall conductivity (Sec.~\ref{sec:hall_con}).
We conclude with a summary and discussion in Sec.~\ref{sec:conc}.

\section{Flux insertion and entanglement} \label{sec:flux}
\subsection{Entanglement as glue}

\begin{figure}[t]
     	\center
	\includegraphics[width=8.0cm]{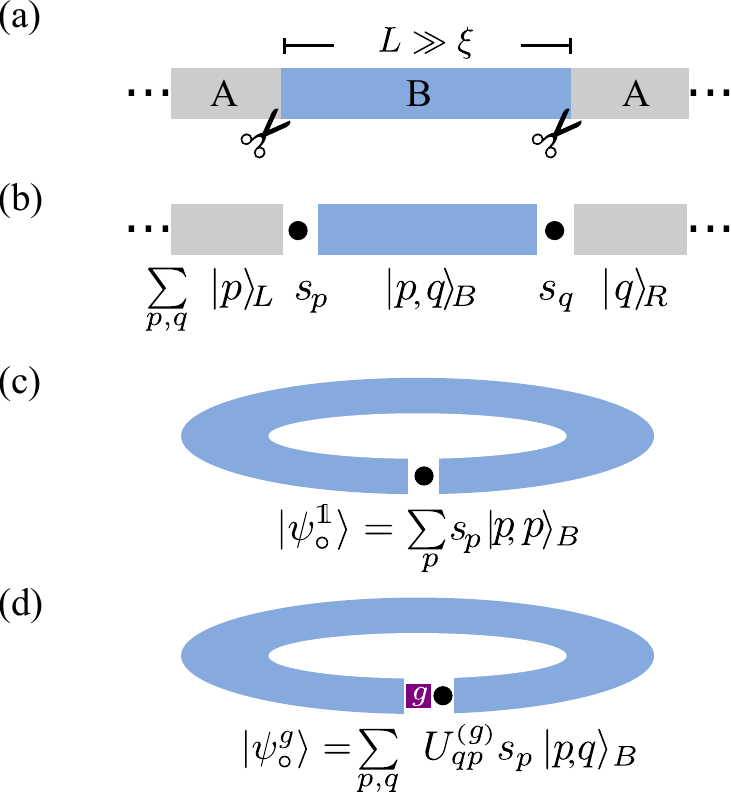}
	\caption{%
		(\textbf{a}) Partition of an infinite quasi-1D system into two pieces, $A$ and $B$. The region $B$ is a finite region of the system, while $A$ extends to infinity in both directions.
		(\textbf{b}) The Schmidt states $B$ of the ground state are decomposed into  tensor products of the left ($p$) and right ($q$) part.
		(\textbf{c}) Tracing over the outer indices of the segment $B$ yields the ground state on a ring.
		(\textbf{d}) In the presence of a symmetry $g$, twisted boundary conditions can be incorporated by acting on the outer indices with a unitary $U^{(g)}$.
	}
	\label{fig:ABA}
\end{figure}

	A gapped ground state of a system with periodic boundary conditions can be obtained by gluing together wave functions for local subsystems.
To be precise, consider a generalized cylinder,  a $d$-dimensional system with geometry $\mathds{R} \times M$. So long as $M$ is finite, we can regard the system to be one-dimensional by viewing $M$  as a single super-site.
Since the system is gapped, the correlation length $\xi$ along the infinite dimension will not depend on the volume of $M$.
Partitioning the system into regions $ABA$, as shown in Fig.~\ref{fig:ABA}(a), we can Schmidt decompose the state as
\begin{equation}
	\ket{\psi} = \sum_a \lambda_a \ket{a}_A \ket{a}_B, \quad \rho_B = \sum_a \lambda_a^2 \ket{a}_B \bra{a}_B.
	\label{eq:SchmidtABA}
\end{equation}
Here $\ket{a}_{A/B}$ are orthogonal basis for $A/B$ and the Schmidt index $a$ labels the quantum fluctuations across the two cuts separating $B$ from $A$. The amplitudes $\lambda_a$ (or the logarithm of them) constitute the `entanglement spectrum.'
For a normalized state, $\sum_a \lambda_a^2 = 1$.

In general the set $\{ \ket{a}_B \}$ forms a complete basis for $B$. However, due to the area law, for a sufficiently large region $B$  only a small subset of states have any significant amplitude $\lambda_a$. In what follows, we will only keep the Schmidt states above some cutoff $\epsilon < \lambda_a$; the cutoff $\epsilon$ can progressively lowered as the size of $B$ increases.

	The crucial point is that if the width of $B$ is large compared to $\xi$, the fluctuations across the left cut should be \emph{independent} from the fluctuations across the right cut. Alternatively, this is the length scale at which the mutual information between the left and right halves of $A$ vanishes, $I(A_L : A_R) \to 0$.
The resulting Schmidt decomposition then has a tensor product structure; labeling the left, right fluctuations by $p$ and $q$, the Schmidt index $a$ can be replaced by the pair $a = (p,q)$.
Using $\lambda_a = s_p s'_q$ and  $\ket{a}_A = \ket{p}_{A_L} \otimes \ket{q}_{A_R}$, we have
\begin{equation}
	\ket{\psi} =  \sum_{p, q} s_p s'_q \ket{p}_{A_L} \ket{p,q}_B \ket{q}_{A_R},
	\label{eq:cuts}
\end{equation}
as illustrated in Fig.~\ref{fig:ABA}(b). We will assume translation invariance, so can set $s_p = s'_p$.

We note that although the set of Schmidt labels $\{a\}$ admits a decomposition into products $\{p\}\times\{q\}$, the wavefunctions in $B$ \emph{do not} decompose as a product: $\ket{a}_B \neq \ket{p}_{B_L} \otimes \ket{q}_{B_R}$.
However, the reduced density matrices for the left/right edges of $B$ depend only on $p, q$ respectively.

An important caveat to Eq.~\eqref{eq:cuts} arises when there is intrinsic topological order, due to the long range entanglement, or in certain topological fermion phases such as the Kitaev chain. \cite{Kitaev-2001, Fidkowski-2010}
Roughly speaking, the left and right cut become correlated with each other by the requirement that identical anyonic flux crosses the two cuts, potentially complicating the tensor product structure $a = p \otimes q$.
Eq.~\eqref{eq:cuts} only applies to a special basis for the degenerate ground states, the `minimally entangled basis', which have definite anyonic flux threading the cylinder \cite{Kitaev-2006, Zhang-2012}.

\subsection{Ground state of a ring}
	Given the decomposition of Eq.~\eqref{eq:cuts}, we can find the wave function for a ring of circumference $L$ for a translational invariant system. This procedure has already been understood in the context of matrix product states \cite{Verstraete-2004, Cincio-2012, ZaletelMongPollmann, Hastings-2013}.
Choosing the two entanglement cuts to be separated by a distance $L$, the ground state of the ring is
\begin{eqnarray}
	\ket{\psi_{\circ}^{(\mathds{1})}} &= \sum_p s_p \ket{p,p}_B + \mathcal{O}(e^{-L/\xi}),
	\label{eq:circwf}
\end{eqnarray}
as shown in Fig.~\ref{fig:ABA}(c). We can show that all the local reduced density matrices are identical to those of the infinite chain, up to exponentially small corrections in $e^{-L/\xi}$. For a local Hamiltonian the energetics of the state are determined by the local density matrices of the system, so reproducing the local density matrices will guarantee a ground state.

	 The interior of each $\ket{p,p}_B$ looks identical to the infinite ground state because the fluctuations remain within a correlation length of the boundaries.
At the edges, note the leftmost region of $\ket{p,q}_B$ is identical to the leftmost region of $\ket{p}_{A_R}$, and similarly for the right.
The fluctuations in Eq.~\eqref{eq:circwf} have been glued together in a manner which is locally identical to the edges of Eq.~\eqref{eq:cuts}, so generate the same reduced density matrices.
A more rigorous justification of Eq.~\eqref{eq:circwf}, and our subsequent results, can be made using matrix product state (MPS) techniques as shown in the Appendix of Ref.~\cite{ZaletelMongPollmann}.

\subsection{Applying a twist}
	Having obtained the ground state for a ring, let us find the ground state when flux under a global symmetry has been threaded through the ring, twisting the boundary condition.\cite{ZaletelMongPollmann, Hastings-2013}
To give a concrete example of `flux', consider a spin system with a $\textrm{U}(1)$ symmetry generated by $S^z$.
The Hamiltonian with flux $\Phi$ has a twist boundary condition between site $1$ and site $L$:
\begin{eqnarray}
	H[\Phi] &= \frac{J}{2} \left [ e^{i \Phi} S_L^+S_{1}^- + e^{- i \Phi} S_L^-S_{L+1}^+ \right ] \nonumber\\
		&\quad + \frac{J}{2} \sum_{i=1}^{L-1} \left [ S_i^+S_{i+1}^-+S_i^-S_{i+1}^+ \right ] \nonumber\\  
		&\quad + \sum_{i=1}^L \left[J S_i^zS_{i+1}^z + D (S_i^z)^2 \right]   \label{eq:spin1}
\end{eqnarray} 
where $J$ is the antiferromagnetic exchange coupling and $D$ represents the single-ion anisotropy parameter. 
We will denote the ground state with flux by $\ket{\psi_{\circ}^{(\Phi)}}$. How should Eq.~\eqref{eq:circwf} be modified to obtain a ground state with flux?

For completeness, we first review the general prescription for inserting flux under any onsite symmetry $g$.
We assume the symmetry is a product of onsite terms, $\hat{g} = \prod_i \hat{g}_i$.
We split the ring into three adjoining regions, $L, R$ and  $E$, so that the symmetry decomposes as $\hat{g} = \hat{g}_L \hat{g}_R \hat{g}_E$.
Since the Hamiltonian is local, it can be decomposed into interactions between the regions
\begin{eqnarray}
	H^{( \mathds{1})} = H_{E L}  + H_{L R} + H_{R E}
	\label{eq:Hnotwist}
\end{eqnarray}
(the intra-region terms can be arbitrarily incorporated into these).
To insert flux $g$, we  twist the boundary condition between region $L, R$ by acting with the symmetry $\hat{g}_L$ (not $\hat{g}_L \hat{g}_R$!) on the interaction $H_{LR}$,
\begin{eqnarray}
	H^{(g)} \equiv H_{E L}  + \hat{g}_L H_{L R} \hat{g}^{-1}_L + H_{R E}.
	\label{eq:Htwist}
\end{eqnarray}
The procedure is unambiguous once the circumference is large compared to the range of the interactions.

To find the ground state of $H^{(g)}$, note that while \emph{globally} $H^{(g)}$ is not unitarily related to $H^{( \mathds{1})}$, \emph{locally} they are.
If $\rho^{(g)}_{LR}$ is the reduced density matrix with flux $g$ for $LR$, we want to minimize the energy for the interactions within subsystem $LR$, 
\begin{eqnarray}
E_{LR} = \mathrm{Tr}( \hat{g}_L H_{L R} \hat{g}^{-1}_L   \rho^{(g)}_{LR} ),
\end{eqnarray}
so want to engineer the twisted density matrix to be $\rho^{(g)}_{LR} = \hat{g}_L \rho^{(\mathds{1})}_{LR} \hat{g}^{-1}_L$.

To do so, return to the Schmidt decomposition of an infinite chain into two half-infinite regions $L, R$: 
\begin{eqnarray}
\ket{\psi} = \sum_p s_p \ket{p}_L \ket{p}_R.
\end{eqnarray}
Since the state is symmetric, the symmetries must act on the left Schmidt states as
\begin{eqnarray}
\hat{g}_L \ket{p}_L = \sum_q  \ket{q}_L U^{(g)}_{q p}
\label{eq:def_U}
\end{eqnarray}
for some matrix $U^{(g)}$ which commutes with $\mathrm{diag}(s)$. The matrices $U^{(g)}$ can be numerically extracted from the ground-state wave function~\cite{Pollmann-2012a}.

\begin{figure}[t]
	\center
	\includegraphics[width=8.0cm]{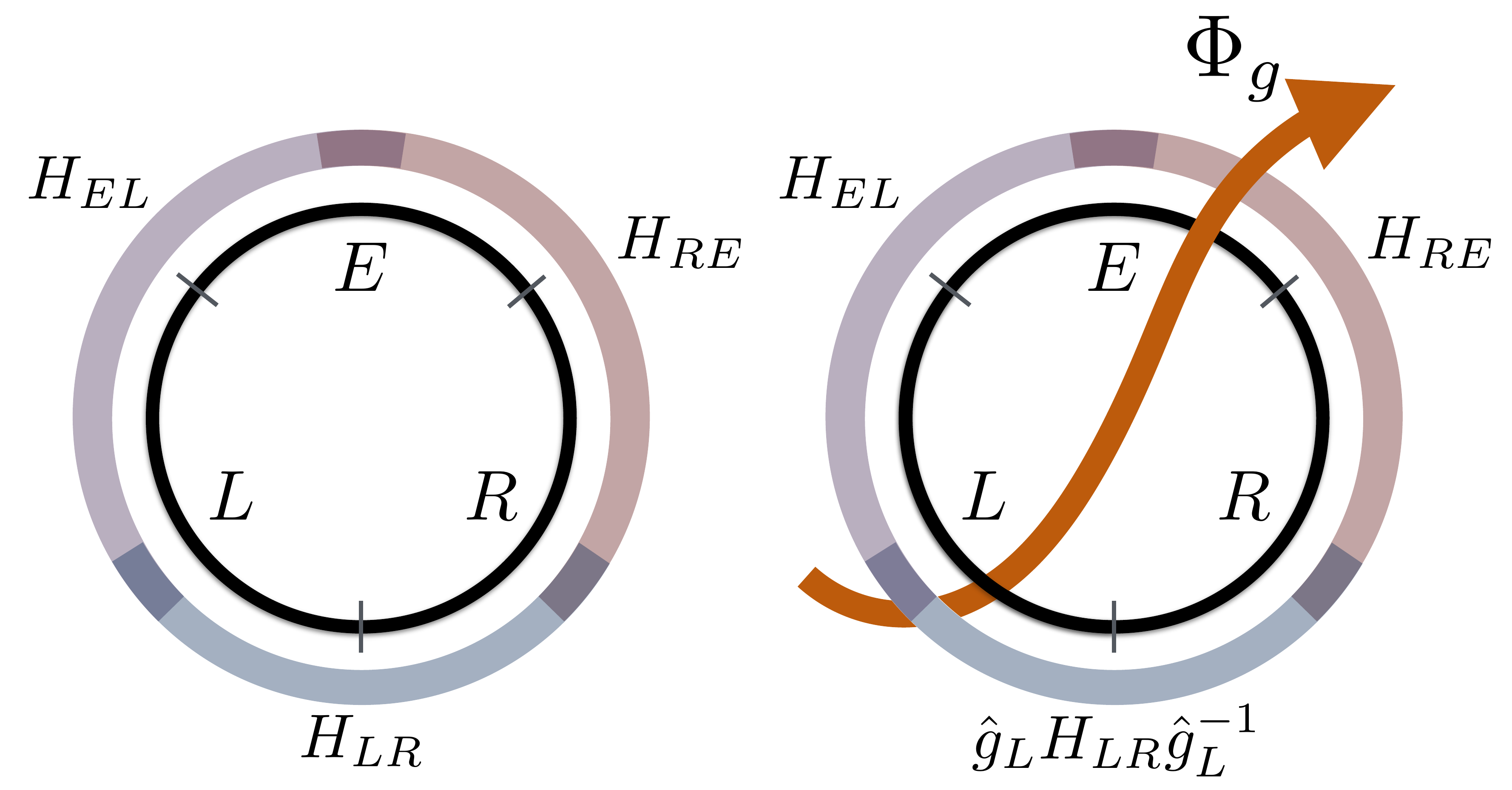}
	\caption{%
		The Hamiltonian for the ring geometry.
		The Hilbert space (inner black ring) is divided into three parts $L$, $R$ and $E$.
		(\textbf{left}) In absence of any flux through the ring, the Hamiltonian [Eq.~\eqref{eq:Hnotwist}] can be decomposed in to three parts; $H_{EL}$, $H_{RE}$, and $H_{LR}$, with their support shown around the black ring.
		Notice that the pieces of the Hamiltonian will have a small overlap with one another.
		(\textbf{right}) With a flux $\Phi_g$ through the ring, the system is described by the Hamiltonian in Eq.~\eqref{eq:Htwist}.
		Conjugating $H_{LR}$ by the symmetry operation $\hat{g}_L$ will only affect terms of the Hamiltonian crossing between regions $L$ and $R$. 
	}
	\label{fig:LRE}
\end{figure}

By definition,
\begin{eqnarray}
\hat{g}_L \ket{\psi} = \sum_{p, q}   \ket{p}_L s_p U^{(g)}_{p q} \ket{q}_R. 
\end{eqnarray}
Such a state has precisely the desired `twisted' density matrix $\rho_{LR}[g]$ in the vicinity of the cut.
So to obtain the twisted ground state on a ring, it is sufficient to insert $U^{(g)}$ into Eq.~\eqref{eq:circwf} before gluing the torus back together, as show in Fig.~\ref{fig:ABA}d): 
\begin{eqnarray}
	\ket{\psi_{\circ}^{(g)}} &= \sum_{p, q} s_p \ket{p,q}_B U^{(g)}_{q p} + \mathcal{O}(e^{-L/\xi}).
	\label{eq:traschmi_flux}
\end{eqnarray}
In conclusion, the ground state of a ring with flux can be obtained from the ground state of the infinite line, Eq.~\eqref{eq:traschmi_flux}, given knowledge of the entanglement spectrum $s_i$ and the action of symmetries on the Schmidt states, $U^{(g)}$.

\begin{figure}[t]
	\center
	\includegraphics[width=8.0cm]{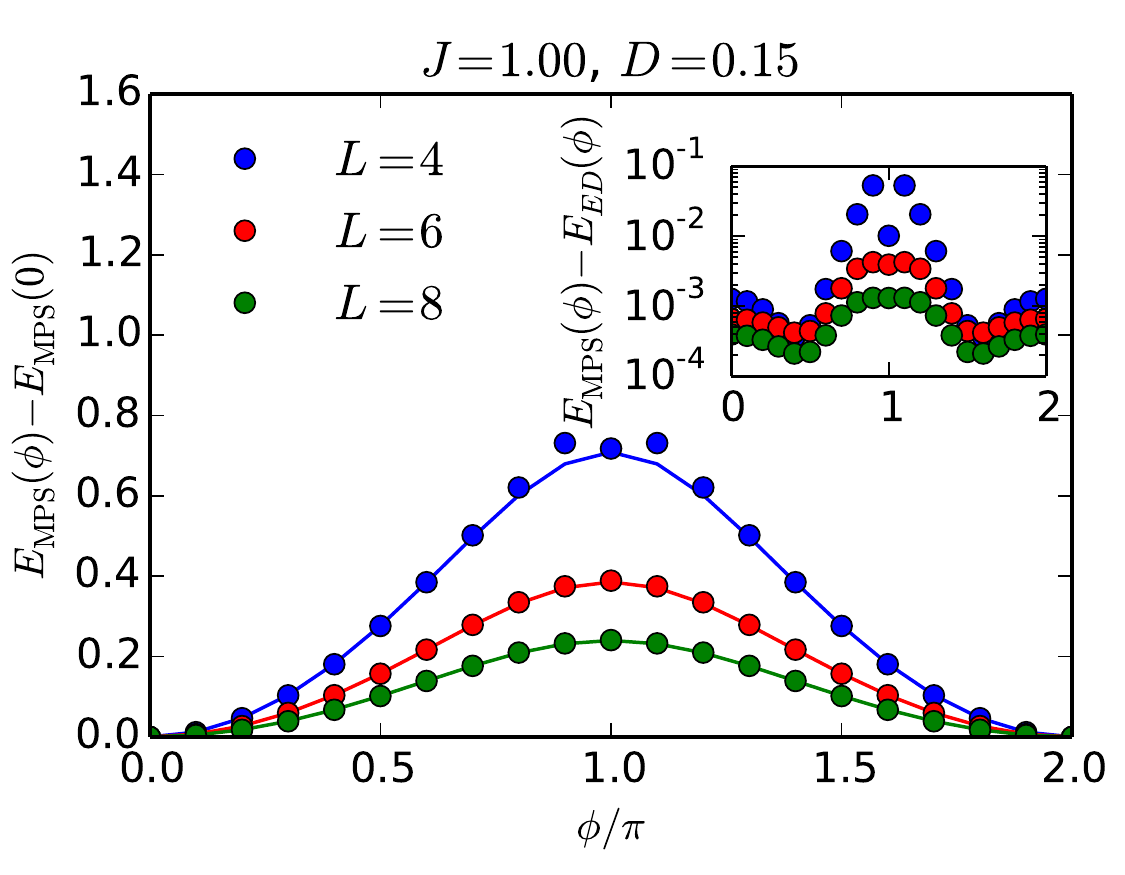}
	\caption{%
		Energies of a spin-1 chain on a ring of length $L$ described by Hamiltonian Eq.~\eqref{eq:spin1} with an inserted flux of $\phi$.
		Energies are obtained both for the wave function $\ket{\psi_{\circ}^{(\Phi)}}$, formula Eq.~\eqref{eq:circwf} (dots), and exactly using diagonalization (line).
		The inset shows the error as compared to the exact ground state wave functions\label{fig:comp}.}
\end{figure}

We numerically test the accuracy of the wave function $\ket{\psi_{\circ}^{(\Phi)}}$ defined in Eq.~\eqref{eq:traschmi_flux} by  comparing with exact diagonalization data for the Hamiltonian of a spin-1 chain defined in Eq.~\eqref{eq:spin1}.
Using the infinite DMRG algorithm in its MPS formulation,\cite{McCulloch-2008, Kjall-2013} it is straightforward to explicitly construct $\ket{\psi_\circ^{(g)}}$.
In particular, we simply calculate the trace over the MPS with inserted boundary operator $U^{(\Phi)}$ to insert the flux.
We test the energetics of the resulting against exact diagonalization, as shown in Fig.~\ref{fig:comp}. 

\subsubsection{Sign-structure for fermions}
\label{sec:fermions}
One additional modification must be made for fermions, due to the Jordan-Wigner string.
The states of a fermionic chain can be expressed through the occupation of bosonic operators $\sigma^+_n = (-1)^{\sum_{j < n} \hat{N}_j }\psi^\dagger_n$, where $\hat{N}_i$ is the fermion occupation at site $i$.
On a periodic chain, the fermionic operators $\psi_n \sim \psi_{n + N}$ are periodic, but because of the string the  $\sigma^+_n$ are not, complicating Eq.~\eqref{eq:traschmi_flux}. 

We restrict ourself to phases in which there are no unpaired Majorana zero-modes at the edge of an open chain. Fermion parity generates a $\mathbb{Z}_2$ symmetry $(-1)^F$, which acts on the left Schmidt states as $U^{(F)}$ [Eq.~\eqref{eq:def_U} for fermion-parity].
For a system with total fermion parity $N_F = 0/1$, careful consideration of the Jordan-Wigner string reveals the ground state of a \emph{periodic} ring in the occupation basis
\begin{eqnarray}
\ket{\psi_{\circ}^{(g)}} &= \sum_{p, q, r} s_p \ket{p q}_B U^{(g)}_{q r} (U^{(F)})^{N_F - 1}_{r p}
	\label{eq:fermionsign}
\end{eqnarray}
Here $U^{(g)}$ generate any other flux we thread through the ring.
The reasoning is that every time a fermion crosses the cut, the Jordan-Wigner string is applied to the other $N_F - 1$ fermions, resulting in a phase; since $U^{(F)}$ changes sign every time a fermion crosses the cut, this factor accounts for the desired sign structure.

	The existence of an emergent Majorana zero-mode at the edge further complicates the analysis, which we discuss briefly in \ref{app:majorana}.

\section{Applications}
\label{sec:examples}

\subsection{Berry phases for U(1) flux insertion}
\label{sec:BerryU1}
In the presence of a continuous symmetry such as $\U1$ charge conservation, we can obtain a Berry phase $\gamma$ for adiabatic flux insertion, known as the Zak-phase in the context of free fermions \cite{Zak-1989}.
It is well known that the Berry phase gives the charge polarization $P$ of the bulk \cite{RestaVanderbilt}.
Intuitively, by threading flux through the ring, we can detect the displacement of charges from their lattice positions.
The polarization $P$ is related to the Berry phase via $P/ae = \gamma/(2\pi)$, where $a$ is lattice spacing and $e$ is the charge quanta.
Notice that the polarization is defined only modulo $ea$ just as the Berry phase is only defined modulo $2\pi$.

Using Eq.~\eqref{eq:traschmi_flux} we show the charge reduces to an average over the entanglement spectrum.
We consider a system on a ring which is invariant under $\U1$ and has a gapped ground state. 
Let us now insert a flux $e^{i \Phi} \in \U1$ through the ring, such that we return to our Hamiltonian at the end of the cycle $\Phi: 0\to 2 \pi$.
As the Hamiltonian $H(\Phi)$ evolves under Eq.~\eqref{eq:Htwist}, so too does its ground state wavefunction $\ket{\psi_\circ^{(\Phi)}}$.
The Berry phase for the process is defined by
\begin{equation}
	\gamma = \int_0^{2\pi}\!\! d\Phi\, A(\Phi) - i \log\braket{\psi_\circ^{(0)}|\psi_\circ^{(2 \pi)}}
\end{equation}
with the Berry connection
\begin{eqnarray}
	A(\Phi) = -i\Braket{ \psi_\circ^{(\Phi)}| \partial_{\Phi} | \psi_\circ^{(\Phi)} } . 
\end{eqnarray}
Using Eq.~\eqref{eq:traschmi_flux}, the ground state with inserted flux is given by $\ket{\psi_\circ^{(\Phi)} } = \sum_{p, q} s_p \ket{p,q}_B U^{(\Phi)}_{q p}$, with $U^{(\Phi)}$ being the representation of the symmetry $e^{i \Phi}$ on the Schmidt states. The Berry connection is thus given by 
\begin{equation}
	A(\Phi) = -i\sum_{p} s_p^2 [(U^{(\Phi)})^{\dag}\partial_{\Phi}U^{(\Phi)}]_{pp}.	\label{eq:A}
\end{equation}
We can always choose a basis such that $U^{(\Phi)} = \mathrm{diag}(e^{i \Phi Q_{p}})$; $Q_p \in \mathbb{Z}$ is the charge of the Schmidt state $\ket{p}_L$.
With this choice, the Berry phase is simply
\begin{eqnarray}
	e^{i\gamma} = \exp\Big[ 2 \pi i\sum_p s^2_p Q_p \Big].
\label{eq:u1_fromQ}
\end{eqnarray}
The desired polarization is  the average charge of the Schmidt state with respect to $s_p^2$.

In 1D, the $\U1$ symmetry alone is not sufficient to guarantee a quantized Berry-phase $\gamma$: there are no non-trivial symmetry protected phases with just $\U1$ in 1D.  But in the presence of an additional symmetry such as time reversal, lattice inversion, or charge-conjugation, the Berry-phase is quantized to $0,\pi$  (this is deeply related to the existence of \emph{symmetry protected topological phases}, see Sec.~\ref{sec:1d_spt} for more details)\cite{Hirano-2008}. 

\paragraph*{Example 1: Affleck Kennedy Lieb Tasaki (AKLT) Chain \cite{AKLT}}. We calculate the Berry phase associated with the adiabatic insertion of a flux of $2\pi$ through the ring \cite{Xu-2008}.
Using Eq.~\eqref{eq:A}, we find
\begin{equation}
	A(\Phi) = \sum_p s^2_p S^z_p,
\end{equation}
where $s_p$ is the Schmidt spectrum and $S^z_p$ is the $S^z$ quantum number of the corresponding Schmidt state. For the AKLT chain, we have two Schmidt states, with $s^2_{p} = \frac12$ for $p = 1,2$.
If we make the assignment $S^z_p = \pm \frac{1}{2}$, we find $\sum_p s^2_p S^a_p = 0$. However, with this choice $\ket{\psi_\circ^{(2 \pi)} } = e^{i \pi} \ket{\psi_\circ^{(0)} }$. Alternatively, we can choose $S^z_1 = 0, S^z_2 = 1$, giving $\sum_p s^2_p S^a_p = \frac{1}{2}$ and $\ket{\psi_\circ^{(2 \pi)} } = \ket{\psi_\circ^{(0)} }$.
In either case it follows that $e^{i\gamma} = -1$. 

\paragraph*{Example 2: Spin-1 Heisenberg chain with single ion anisotropy.}
\begin{figure}[t]
	\centering
	\includegraphics[width=8.0cm]{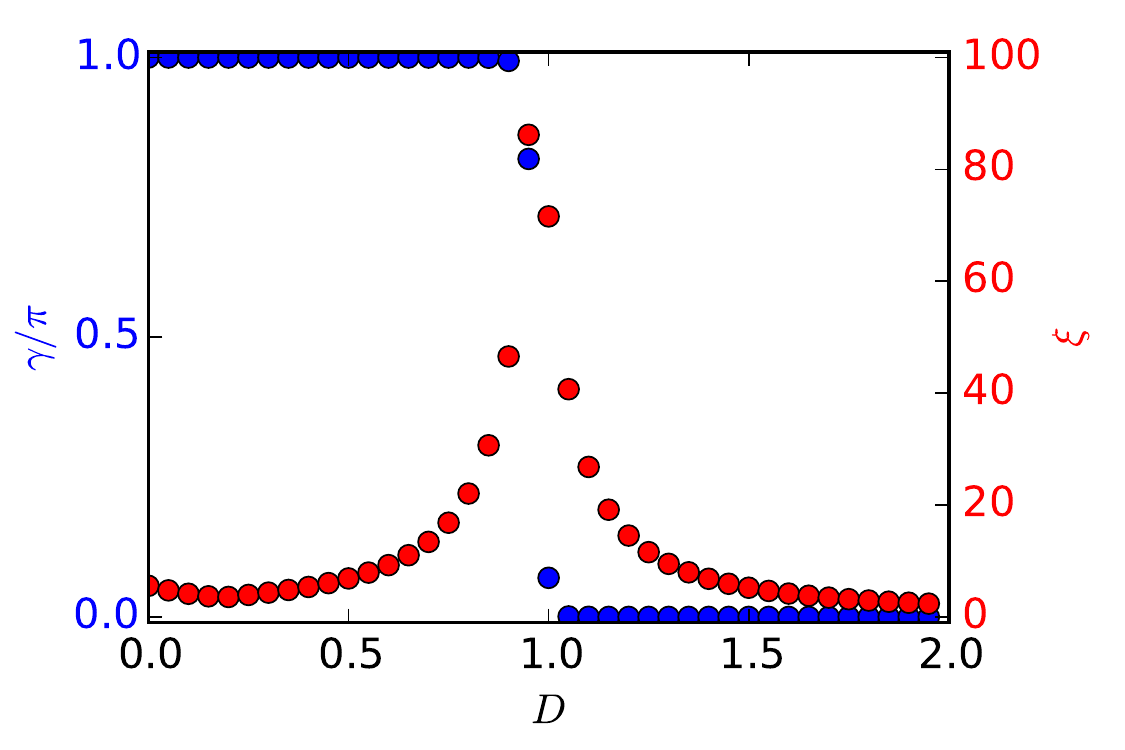}
	\caption{%
		Flux insertion for the Heisenberg model with single-ion anisotropy. The blue dots show the Berry phase associated with the adiabatic flux insertion of $2\pi$, distinguishing the Haldane phase ($\gamma=\pi$) and the large-$D$ phase ($\gamma=0$).
		The red dots show the correlation length which diverges at $D\approx1$.}
\label{fig:haldane}
\end{figure}
We consider Hamiltonian  Eq.~(\ref{eq:spin1}) and use the parameter $D$ to tune across a phase transition between a non-trivial Haldane phase and a trivial phase. We then  attempt to detect with the $\U1$ Berry phase. The entanglement spectrum and the quantum numbers are  obtained using iDMRG, in order to calculate $e^{i \gamma}$ using Eq.~\eqref{eq:u1_fromQ}. The results are shown in Fig.~\ref{fig:haldane}, and indeed gives a precise determination of the phase transition. The same result had been obtained in Ref.~\cite{Hirano-2008} by explicitly obtaining the ground states with inserted fluxes using exact diagonalization.

\paragraph*{Example 3: The modular $T$-transformation.}
	Finally, we consider a 2D state on a torus, obtained from the plane by identifying points $(x, y) \sim (x, y + L_y)$, $(x + L_x, y) \sim (x, y + \tau_r L_y)$.
The parameter $\tau = \tau_r + iL_x/L_y$ is called the modular parameter.
Since $\tau \to \tau + 1$ defines an identical torus, we can calculate a Berry-phase $U_T$ as the ground state $\ket{\tau} \to \ket{\tau + 1}$.
Known as the modular $T$-matrix, $U_T$ encodes the topological spin of the anyons and chiral central charge of the edge, both of which are topological invariants of the state \cite{Keski-Vakkuri-1993}. To relate this to a $\U1$ Berry phase, we view the torus to be a continuous 1D system by collapsing each ring at fixed $x$ to a single `site.'
The translation $e^{i \hat{P}^y \Delta y}: y \to y + \Delta y$ is then an \emph{onsite} $\U1$ symmetry of the system.
In this language, the modular parameter $\tau_r$ is nothing but $\U1$ flux under translation: $\Phi_{\hat{P}^y} \equiv 2 \pi \tau_r$.
Assigning $y$-momenta $k_p \in \frac{2 \pi}{L_y} \mathbb{Z}$ to each of the Schmidt states, we immediately conclude the $T$-matrix is
\begin{eqnarray}
	U_T = e^{i \sum_p s_p^2 k_p L_y }.
\end{eqnarray}
Hence the chiral central charge and topological spin of the anyons are trivially encoded in the entanglement spectrum.
This procedure has been successfully carried out for various quantum Hall systems, \cite{ZaletelMongPollmann} topological lattice models, \cite{TuZhangQi} and 2D SPT phases \cite{Zaletel-2013}.

\subsection{Response theory of 1D symmetry protected topological phases}
\label{sec:1d_spt}
In the presence of symmetries, 1D systems have a rich variety of phases generalizing the physics of the spin-1 Haldane chain: symmetry protected topological (SPT) phases \cite{Gu-2009, Pollmann-2010, Fidkowski-2010, Pollmann-2012,Chen-2011,Schuch-2011}.  So long as the Hamiltonian preserves the symmetry group $G$, the ground state of an SPT phase is distinct from a trivial product state.
The essence of an SPT state is that an open chain has edge states which transform under a \emph{projective} representation of the symmetry group $G$.
Unlike an ordinary representation, projective representations may carry additional phase factors under composition;
i.e., if  $g, h \in G$, then 
\begin{eqnarray}
	V^{(g)} V^{(h)} = \omega(g, h) V^{(gh)}
\end{eqnarray}
where the set of complex numbers $\omega(g, h) \in \U1$ is called the ``factor set''.
(Were $V$'s to form a regular representation of $G$, then all the $\omega$ would equal to 1.)

For each projective symmetry representation of $G$, the set of $\omega$ is not unique: one can always construct a new set $\omega'$ from $\omega$ by multiplying each $V$ by some arbitrary phase.
For example, by letting $V^{(g)} \mapsto e^{i\theta_g}V^{(g)}$, the factor set transform as $\omega(g, h) \mapsto e^{i(\theta_g + \theta_h - \theta_{gh})} \omega(g, h)$.
The allowed factor sets, modulo the above phase transform, define distinct projective representations, leading to a classification of 1D bosonic and fermionic phases with symmetry \cite{Gu-2009, Pollmann-2010, Fidkowski-2010, Pollmann-2012,Chen-2011,Schuch-2011}. 
Certain ratios, such as $\omega(g, h) / \omega(h, g)$ when $g$ and $h$ commute, are invariants of the projective representation so are candidate  physical observables.

	There is a deep relationship between the symmetry transformations of the edge states $V^{(g)}$ and the matrices $U^{(g)}$ which encode how the Schmidt states transform under the symmetry, Eq.~\eqref{eq:def_U}.
While the matrices themselves needn't be the same (there will generically be many more Schmidt states then edge states), the \emph{factor set} $\omega(g, h)$ must be identical, modulo the discussed phase ambiguity.
This gives an entanglement point of view on SPT phases: the Schmidt states of an SPT state transform \emph{projectively} under the symmetry group. In Ref.~\cite{Pollmann-2012a}, this was used to give a numerical procedure to complete characterize an 1D SPT state given its ground state.

	It is useful to have further physical signatures of the topological order. In fact, a 1D SPT on a ring has a quantized response when flux threads a ring\cite{Kitaev-2001, WenMonodromy-2013}.
Let $\ket{\psi_\circ^{(\mathds{1})}}$ denote the ground state of a ring with periodic boundary conditions, which for simplicity we assume is neutral under all symmetries: $\hat{h} \ket{\psi_\circ^{(\mathds{1})}} = \ket{\psi_\circ^{(\mathds{1})}}$ (otherwise all statements below must be interpreted relative to the charge of the ground state).
We then thread flux under a symmetry $g$ to obtain a new ground state $\ket{\psi_\circ^{(g)}}$, and measure the charge of the state under another symmetry $h$ which commutes with $g$:
\begin{eqnarray}
i_{g}(h) \ket{\psi_\circ^{(g)}} \equiv \hat{h} \ket{\psi_\circ^{(g)}}.
\end{eqnarray}
For a trivial phase, no charge is induced ($i_{g}(h) = 1$), while for an SPT phase $g$-flux can induces a $h$-charge ($i_{g}(h) \neq 1$), giving  a physical probe of the order.
	
	We now use the flux-insertion trick to give a simple proof of this statement---providing a direct link between the symmetry properties of the entanglement spectrum and the physical observable $i_{g}(h)$.
Recall that for each symmetry $g \in G$ we compute from the ground state the matrices $U^{(g)}$ which encode how the symmetry acts on the left Schmidt states, shown in Eq.~\eqref{eq:def_U}.
As discussed, the Schmidt states may transform projectively, meaning
\begin{eqnarray}
	U^{(g)} U^{(h)} = \omega(g, h) U^{(gh)}.
	\label{eq:proj_U}
\end{eqnarray}
%We will combine the flux-threading trick with Eq. \eqref{eq:prom_U} to derive the physical response of the SPT.
Using the flux-threading trick, we can write the ground state of the ring with flux $\ket{\psi_\circ^{(g)}}$ in terms of $\ket{p,q}_B$ and $U^{(g)}$.

We denote $\ket{B}$ to be a matrix of wavefunctions $\ket{p,q}_B$, and $s$ to be a diagonal matrix with elements $s_p$ along the diagonal.
Hence a wavefunction on a ring can be expressed succinctly as
\begin{eqnarray}
	\ket{\psi_\circ^{(\mathds{1})}} = \mathrm{Tr}[\ket{B} s] .
\end{eqnarray}
With the insertion of a flux, we have $\ket{\psi_\circ^{(g)}} = \mathrm{Tr}[\ket{B} U^{(g)} s]$.
It is important to note that $s$ commutes with $U^{(g)}$; this is a statement that Schmidt states within an irreducible representation must have the same Schmidt weight \cite{Pollmann2010}.
Finally, we also make use of how $\ket{B}$ transform under symmetry $h$: $\hat{h} \ket{B} = {U^{(h)}}^{-1} \ket{B} U^{(h)}$.
We can then measure the charge of the resulting state:
\begin{eqnarray}
	\hat{h} \ket{\psi_\circ^{(g)}}
	&= \mathrm{Tr} \big[ \hat{h} \ket{B} U^{(g)} s \big] \nonumber\\
	&= \mathrm{Tr} \big[ {U^{(h)}}^{-1} \ket{B} U^{(h)} U^{(g)} s \big]\nonumber  \\
	&= \omega(h, g) \mathrm{Tr} \big[ {U^{(h)}}^{-1} \ket{B} U^{(hg)} s \big] \nonumber \\
	&= \omega(h, g) \mathrm{Tr} \big[ {U^{(h)}}^{-1} \ket{B} U^{(gh)} s \big]  \\
	&= \frac{\omega(h, g)}{\omega(g, h)} \mathrm{Tr} \big[ {U^{(h)}}^{-1} \ket{B} U^{(g)} U^{(h)} s \big] \nonumber\\
	&= \frac{\omega(h, g)}{\omega(g, h)} \mathrm{Tr} \big[ {U^{(h)}}^{-1} \ket{B} U^{(g)} s \, U^{(h)} \big] \nonumber\\
	&= \frac{\omega(h, g)}{\omega(g, h)} \mathrm{Tr} \big[ \ket{B} U^{(g)} s \big] \nonumber\\
	&= \frac{\omega(h, g)}{\omega(g, h)} \ket{\psi_\circ^{(g)}} \equiv i_{g}(h) \ket{\psi_\circ^{(g)}} \nonumber
\end{eqnarray}
Hence $i_{g}(h) = \frac{\omega(h, g)}{\omega(g, h)}$.
Therefore we've proven the factor set of the projective representation encodes the charge induced by flux insertion.

\subsection{Hall conductance}
\label{sec:hall_con}
	Finally, we consider how to efficiently calculate the Hall conductance $\sigma^{xy}$ of an interacting 2D system using the entanglement spectrum of a cylinder, the geometry relevant to finite and infinite DMRG.
Equivalent physics has been discussed in Ref.~\cite{Alexandradinata-2011} as `spectral flow' in the entanglement Hamiltonian.
To review, on a torus the Hall conductance can be obtained by finding  many-body ground states $\ket{\Phi_x, \Phi_y}_\textrm{torus}$ for a torus in which fluxes $\Phi_{x/y}$ are thread through the two cycles of the torus \cite{NiuThoulessWu-1985}.
By repeatedly solving for the ground state in a 2D discrete grid of $\Phi_i$, we obtain the discretized Berry-connection $\mathcal{A} = \bra{\Phi_x, \Phi_y} i \nabla \ket{\Phi_x, \Phi_y}_\textrm{torus}$.
The Chern number 
\begin{eqnarray}
	\frac{e^2}{2\pi h} \int d\Phi_x d\Phi_y \, \nabla \times \mathcal{A}(\Phi_x, \Phi_y) 
\end{eqnarray}
is the desired Hall conductance.

\begin{figure}[t]
	\centering
	\includegraphics[width=5.0cm]{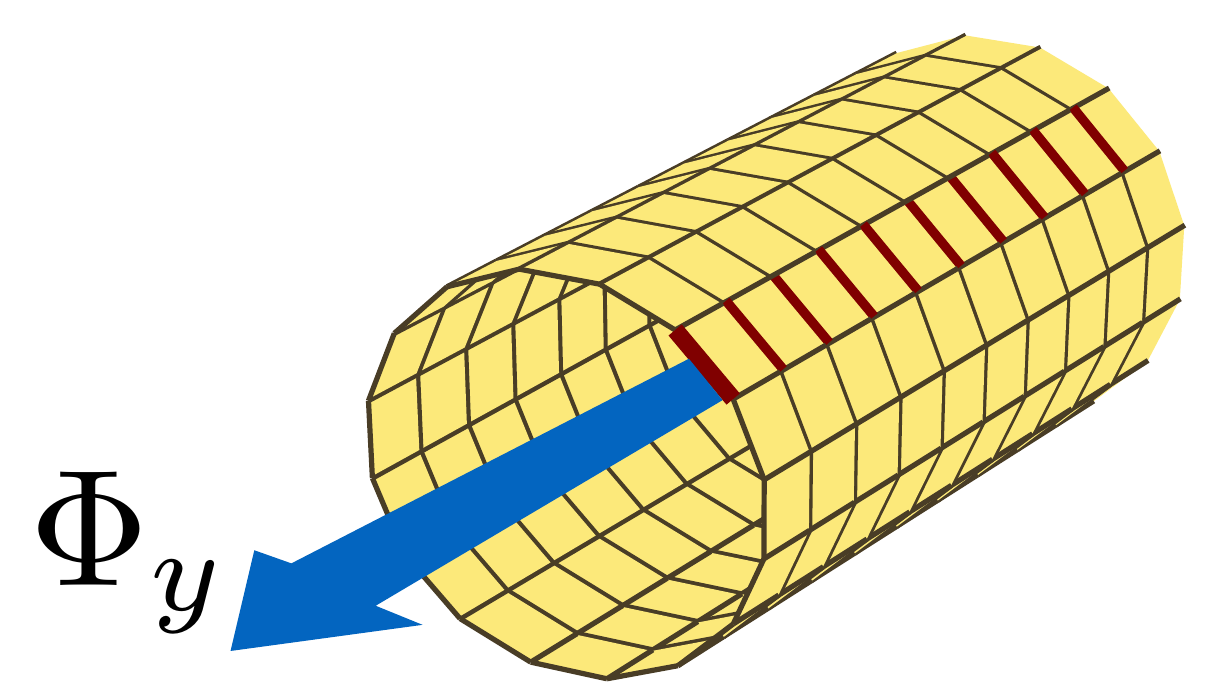}
	\caption{%
		Measuring Hall conductance of a phase by solving for the ground state on a cylinder for various flux $\Phi_y$.
		The $y$-flux through the cylinder is implemented by altering terms of the Hamiltonian ($a_{ij} = 1$) along a single row of bonds, colored red in the figure.  (We assume nearest-neighbor interactions for the purpose of illustration.)
		For each value of $\Phi_y$, we measure the charge polarization of the cylinder via an entanglement cut aronud the cylinder.
	}
	\label{fig:cylflux}
\end{figure}

	However, we know the entanglement spectrum of a cylinder allows us to simulate one direction of flux insertion---say $\Phi_x$---for free.
This can be used to greatly reduce the number of ground states required to calculate the Hall conductance. For concreteness, consider a bosonic or fermionic tight binding model on a cylinder with periodic coordinate $y$,
\begin{eqnarray}
	H[\Phi_y] = -\sum_{\mu, \nu} e^{i \Phi_y a_{\mu\nu} } t_{\mu\nu} + \sum_{\mu, \nu} U_{\mu\nu} n_\mu n_\nu.
\end{eqnarray}	
The indices $\mu$, $\nu$ denotes sites of the lattice, and may include additional bands, spins, etc.
To generate flux $\Phi_y$ through the cylinder, we modulate the hoppings $t_{\mu\nu}$ by a phase $e^{i \Phi_y a_{\mu\nu} }$, where $a_{\mu\nu} = 1$ for all bonds which cross a cut at $y = 0$, and $a_{\mu\nu} = 0$ elsewhere.
The relevant bonds are illustrated in Fig.~\ref{fig:cylflux}.
	
	We start by obtaining the ground state of $H[\Phi_y]$  using the iDMRG method, which we denote by $\ket{\Phi_y}_\textrm{cyl}$.
Fixing $\Phi_y$, we can calculate the $\U1$ Berry-phase for flux insertion $\Phi_x$ using only the entanglement spectrum, as discussed in Sec.~\ref{sec:BerryU1}.
From the entanglement spectrum $\{s_p, Q_p\}$,  with Schmidt values $s_p$ and the $\U1$ quantum numbers $Q_p \in \mathbb{Z}$, the $\U1$ Berry-phase is, analogous to Eq.~(\ref{eq:u1_fromQ}),
\begin{equation}
	e^{i \gamma(\Phi_y)} = \exp\big[ 2 \pi i\sum_p (s_p^{\Phi_y})^2 Q^{\Phi_y}_p \big].
\label{eq:hall_gamma}
\end{equation}
We emphasize Eq.~\eqref{eq:hall_gamma} holds for the entanglement spectrum of both finite length and infinite cylinders.
The 2D Hall conductance is
\begin{eqnarray}
	\sigma^{xy} &= \frac{e^2}{2\pi h} \int d\Phi_x d\Phi_y \, \nabla \times \mathcal{A}(\Phi_x, \Phi_y)	\\
			    &= \frac{e^2}{2\pi h} \int_0^{2\pi}\!\!\! d\Phi_y \, \partial_{\Phi_y} \gamma(\Phi_y) .
\end{eqnarray}

Hence, using only a coarse set of cylinder ground states $\ket{\Phi_y}_\textrm{cyl}$, we obtain the Hall conductance.
For tensor network methods this is generally significantly easier then finding a 2D grid of torus ground states $\ket{\Phi_x, \Phi_y}_\textrm{torus}$.

\section{Conclusions and discussions}
\label{sec:conc}
Equation \eqref{eq:traschmi_flux} allows us to obtain the ground state of a ring with arbitrary flux given the ground state of an infinite chain, the entanglement spectrum $s_a$, and knowledge of how symmetries act on Schmidt states, $U^{(g)}$. 
Our main interest in this result arises because numerous topological invariants associated with symmetries, such as the Hall conductance and topological spin, can be computed via the response to flux insertion through a cycle of the system.
For instance, the Hall conductance can be computed by the charge pumped around one cycle of a torus as flux is threaded through the other.
Eq.~\eqref{eq:traschmi_flux} demonstrates that these responses are in fact encoded in the entanglement structure of the \emph{infinite} system.

\section{Acknowledgements}
MZ would like to thank Joel Moore and support from NSF DMR-1206515.
RM is supported by the Sherman Fairchild Foundation.
MZ and RM acknowledges the hospitality of the Dresden visitor program.

\appendix

\section{Wavefunctions for topological fermion phases}
\label{app:majorana}
When the system has emergent Majorana zero-modes at the boundary, such as in the Kitaev-chain, the tensor product structure of Eq.~\eqref{eq:cuts} isn't strictly correct.
A detailed discussion of the entanglement structure in this case is given in Refs.~\cite{Fidkowski-2010,Turner-2010}.
Additional subtleties arise when we use a bosonic representation for fermions; the Jordan-Wigner string behaves non-trivially when placed on a ring.

The bipartite Schmidt decomposition $ABA$ [Eq.~\eqref{eq:SchmidtABA}] can be chosen such that each Schmidt state $\ket{a}_B$ is an eigenvector of fermion parity.
Thus the set of Schmidt labels $\{a\}$ may be partitioned into two subsets, $\{a\}_{\pm}$  according to the fermion parity $\pm$ of $\ket{a}_B$.
When the system has a Majorana zero-mode, the labels $\{a\}$ don't have a tensor product structure. However, each parity sector individually does: $\{a\}_{\pm} \sim \{ (p, q)\}_{\pm}$.
Thus, we can build two wavefunctions on a ring, distinguished by their fermion parity:
\begin{eqnarray}\label{eq:kitaev_fermionsign}
	\ket{\psi_\circ^+} &= \sum_{p \mid (p,p) \in +} s_p \ket{p, p}_B ,
\\	\ket{\psi_\circ^-} &= \sum_{p \mid (p,p) \in -} s_p \ket{p, p}_B .
\end{eqnarray}
In the bosonic occupation language, where fermion operators are accompanied by a Jordan-Wigner string, both of these states have periodic boundary condition.
However, as discussed in Sec.~\ref{sec:fermions}, the map between bosonic degrees of freedom to fermions alters the boundary condition.
\begin{eqnarray}
	\Phi_\textrm{boson} = - (-1)^\textrm{\# fermions} \Phi_\textrm{fermion} .
\end{eqnarray}
Thus the two states in Eq.~\eqref{eq:kitaev_fermionsign} have differing boundary conditions as a fermionic system. 
One identifying characteristic of the Kitaev-chain state is that the fermion parity of the ground state differs between periodic (no flux) and antiperiodic ($\pi$ flux) boundary conditions.
This aspect is automatically captured in the construction of the wavefunctions on a ring.

\section*{References}

%\bibliographystyle{unsrt}
%\bibliography{topo.bib}

\begin{thebibliography}{10}

\bibitem{Wen-1990}
Xiao-Gang Wen.
\newblock {\em Int. J. Mod. Phys.}, B4:239, 1990.

\bibitem{Levin-2006}
Michael Levin and Xiao-Gang Wen.
\newblock Detecting topological order in a ground state wave function.
\newblock {\em Phys. Rev. Lett.}, 96(11):110405, 2006.

\bibitem{Kitaev-2006}
Alexei Kitaev and John Preskill.
\newblock Topological entanglement entropy.
\newblock {\em Phys. Rev. Lett.}, 96(11):110404, 2006.

\bibitem{White-1992}
Steven~R. White.
\newblock Density matrix formulation for quantum renormalization groups.
\newblock {\em Phys. Rev. Lett.}, 69(19):2863--2866, Nov 1992.

\bibitem{Li-2008}
Hui Li and F.~D.~M. Haldane.
\newblock Entanglement spectrum as a generalization of entanglement entropy:
  Identification of topological order in non-abelian fractional quantum hall
  effect states.
\newblock {\em Phys. Rev. Lett.}, 101:010504, Jul 2008.

\bibitem{Papic-2011}
Z.~Papi\ifmmode~\acute{c}\else \'{c}\fi{}, B.~A. Bernevig, and N.~Regnault.
\newblock Topological entanglement in abelian and non-abelian excitation
  eigenstates.
\newblock {\em Phys. Rev. Lett.}, 106:056801, Feb 2011.

\bibitem{Chandran-2014}
A.~{Chandran}, V.~{Khemani}, and S.~L. {Sondhi}.
\newblock {\em ArXiv e-prints}, November 2013.

\bibitem{Zak-1989}
J.~Zak.
\newblock BerryÕs phase for energy bands in solids.
\newblock {\em Phys. Rev. Lett.}, 62:2747--2750, Jun 1989.

\bibitem{RestaVanderbilt}
Raffaele Resta and David Vanderbilt.
\newblock Theory of polarization: A modern approach.
\newblock In {\em Physics of Ferroelectrics}, volume 105 of {\em Topics in
  Applied Physics}, pages 31--68. Springer Berlin Heidelberg, 2007.

\bibitem{Laughlin-1981}
R.~B. Laughlin.
\newblock Quantized hall conductivity in two dimensions.
\newblock {\em Phys. Rev. B}, 23:5632--5633, May 1981.

\bibitem{NiuThoulessWu-1985}
Qian Niu, D.~J. Thouless, and Yong-Shi Wu.
\newblock Quantized hall conductance as a topological invariant.
\newblock {\em Phys. Rev. B}, 31:3372--3377, Mar 1985.

\bibitem{Alexandradinata-2011}
A.~Alexandradinata, Taylor~L. Hughes, and B.~Andrei Bernevig.
\newblock Trace index and spectral flow in the entanglement spectrum of
  topological insulators.
\newblock {\em Phys. Rev. B}, 84:195103, Nov 2011.

\bibitem{Keski-Vakkuri-1993}
Esko Keski-Vakkuri and Xiao-Gang Wen.
\newblock {The Ground State Structure and Modular Transformations of Fractional
  Quantum Hall States on a Torus}.
\newblock {\em International Journal of Modern Physics B}, 07(25):4227--4259,
  1993.

\bibitem{WenMonodromy-2013}
X.-G. {Wen}.
\newblock {\em ArXiv e-prints}, January 2013.

\bibitem{Kitaev-2001}
A~Kitaev.
\newblock Unpaired majorana fermions in quantum wires.
\newblock {\em Physics-Uspekhi}, 44(10S):131, 2001.

\bibitem{Fidkowski-2010}
Lukasz Fidkowski and Alexei Kitaev.
\newblock Effects of interactions on the topological classification of free
  fermion systems.
\newblock {\em Phys. Rev. B}, 81(13):134509, Apr 2010.

\bibitem{Zhang-2012}
Yi~Zhang, Tarun Grover, Ari Turner, Masaki Oshikawa, and Ashvin Vishwanath.
\newblock Quasiparticle statistics and braiding from ground-state entanglement.
\newblock {\em Phys. Rev. B}, 85:235151, Jun 2012.

\bibitem{Verstraete-2004}
F.~Verstraete, D.~Porras, and J.~I. Cirac.
\newblock Density matrix renormalization group and periodic boundary
  conditions: A quantum information perspective.
\newblock {\em Phys. Rev. Lett.}, 93:227205, Nov 2004.

\bibitem{Cincio-2012}
L.~Cincio and G.~Vidal.
\newblock Characterizing topological order by studying the ground states on an
  infinite cylinder.
\newblock {\em Phys. Rev. Lett.}, 110:067208, Feb 2013.

\bibitem{ZaletelMongPollmann}
Michael~P. Zaletel, Roger S.~K. Mong, and Frank Pollmann.
\newblock Topological characterization of fractional quantum hall ground states
  from microscopic hamiltonians.
\newblock {\em Phys. Rev. Lett.}, 110:236801, Jun 2013.

\bibitem{Hastings-2013}
M.~B. Hastings.
\newblock Classifying quantum phases with the kirby torus trick.
\newblock {\em Phys. Rev. B}, 88:165114, Oct 2013.

\bibitem{Pollmann-2012a}
Frank Pollmann and Ari~M. Turner.
\newblock Detection of symmetry-protected topological phases in one dimension.
\newblock {\em Phys. Rev. B}, 86:125441, Sep 2012.

\bibitem{McCulloch-2008}
I.~P. McCulloch.
\newblock {Infinite size density matrix renormalization group, revisited}.
\newblock Unpublished, 2008.

\bibitem{Kjall-2013}
Jonas~A. Kj\"all, Michael~P. Zaletel, Roger S.~K. Mong, Jens~H. Bardarson, and
  Frank Pollmann.
\newblock Phase diagram of the anisotropic spin-2 xxz model: Infinite-system
  density matrix renormalization group study.
\newblock {\em Phys. Rev. B}, 87:235106, Jun 2013.

\bibitem{Hirano-2008}
T.~Hirano, H.~Katsura, and Y.~Hatsugai.
\newblock Topological classification of gapped spin chains: Quantized berry
  phase as a local order parameter.
\newblock {\em Phys. Rev. B}, 77(9):094431, 2008.

\bibitem{AKLT}
I.~Affleck, T.~Kennedy, E.~Lieh, and H.~Tasaki.
\newblock {\em Phys. Rev. Lett.}, 59:799, 1988.

\bibitem{Xu-2008}
Ying Xu, Hosho Katsura, Takaaki Hirano, and VladimirE. Korepin.
\newblock Entanglement and density matrix of a block of spins in aklt model.
\newblock {\em Journal of Statistical Physics}, 133(2):347--377, 2008.

\bibitem{TuZhangQi}
Hong-Hao Tu, Yi~Zhang, and Xiao-Liang Qi.
\newblock Momentum polarization: An entanglement measure of topological spin
  and chiral central charge.
\newblock {\em Phys. Rev. B}, 88:195412, Nov 2013.

\bibitem{Zaletel-2013}
M.~P. {Zaletel}.
\newblock {Detecting two dimensional symmetry protected topological order in a
  ground state wave function}.
\newblock unpublished, September 2013.

\bibitem{Gu-2009}
Z-C. Gu and X-G. Wen.
\newblock Tensor-entanglement-filtering renormalization approach and symmetry
  protected topological order.
\newblock {\em Phys. Rev. B}, 80:155131, 2009.

\bibitem{Pollmann-2010}
Frank Pollmann, Ari~M. Turner, Erez Berg, and Masaki Oshikawa.
\newblock Entanglement spectrum of a topological phase in one dimension.
\newblock {\em Phys. Rev. B}, 81(6):064439, Feb 2010.

\bibitem{Pollmann-2012}
Frank Pollmann, Erez Berg, Ari~M. Turner, and Masaki Oshikawa.
\newblock Symmetry protection of topological phases in one-dimensional quantum
  spin systems.
\newblock {\em Phys. Rev. B}, 85:075125, Feb 2012.

\bibitem{Chen-2011}
Xie Chen, Zheng-Cheng Gu, and Xiao-Gang Wen.
\newblock Classification of gapped symmetric phases in one-dimensional spin
  systems.
\newblock {\em Phys. Rev. B}, 83:035107, Jan 2011.

\bibitem{Schuch-2011}
{Schuch, N. and P\'erez-Garc\'ia, D. and Cirac, J. I.}
\newblock Classifying quantum phases using matrix product states and projected
  entangled pair states.
\newblock {\em Phys. Rev. B}, 84:165139, Oct 2011.

\bibitem{Pollmann2010}
Frank Pollmann, Ari~M. Turner, Erez Berg, and Masaki Oshikawa.
\newblock Entanglement spectrum of a topological phase in one dimension.
\newblock {\em Phys. Rev. B}, 81:064439, Feb 2010.

\bibitem{Turner-2010}
Ari~M. Turner, Frank Pollmann, and Erez Berg.
\newblock Topological phases of one-dimensional fermions: An entanglement point
  of view.
\newblock {\em Phys. Rev. B}, 83:075102, Feb 2011.

\end{thebibliography}

\end{document}